\documentclass[prl,showpacs,preprintnumbers,amsmath,amssymb,twocolumn]{revtex4-1}
\usepackage{graphicx}
\usepackage{dcolumn}
\usepackage{bm}
\usepackage{hyperref}
\usepackage{xcolor}
\hypersetup{
colorlinks,
linkcolor={blue!100!black},
citecolor={blue!50!black},
urlcolor={blue!80!black}}
\begin{document}
\title{Directional Phonon Suppression Function as a Tool for the Identification of Ultralow Thermal Conductivity Materials}
\author{Giuseppe Romano}
\affiliation{Department of Mechanical Engineering, Massachusetts Institute of Technology, 77 Massachusetts Avenue, Cambridge, MA 02139}
\email{romanog@mit.edu}
\author{Alexie M. Kolpak}
\affiliation{Department of Mechanical Engineering, Massachusetts Institute of Technology, 77 Massachusetts Avenue, Cambridge, MA 02139}
\email{kolpak@mit.edu}
\begin{abstract}
Boundary-engineering in nanostructures has the potential to dramatically impact the development of materials for high-efficiency conversion of thermal energy directly into electricity. In particular, nanostructuring of semiconductors can lead to strong suppression of heat transport with little degradation of electrical conductivity. Although this combination of material properties is promising for thermoelectric materials, it remains largely unexplored. In this work, we introduce a novel concept, the directional phonon suppression function, to unravel boundary-dominated heat transport in unprecedented detail. Using a combination of density functional theory and the Boltzmann transport equation, we compute this quantity for nanoporous silicon materials. We first compute the thermal conductivity for the case with aligned circular pores, confirming a significant thermal transport degradation with respect to the bulk. Then, by analyzing the information on the directionality of phonon suppression in this system, we identify a new structure of rectangular pores with the same porosity that enables a four-fold decrease in thermal transport with respect to the circular pores. Our results illustrate the utility of the directional phonon suppression function, enabling new avenues for systematic thermal conductivity minimization and potentially accelerating the engineering of next-generation thermoelectric devices.\end{abstract}
\maketitle
\today 
\section{Introduction}
Understanding heat transport in the presence of nanoscale boundaries is of paramount significance for many applications, including thermoelectrics~\cite{Rowe1995-po,Vineis2010-lk}, thermal rectification~\cite{Li2004-ha} and thermal dissipators~\cite{Lai1996-eg}. When the dominant phonon mean free path (MFP) approaches the characteristic length scale of a material, classical phonon size effects take place, leading to a decrease in thermal transport~\cite{Chen2005-ac}. For example, first-principles calculations of silicon show that half of the heat is carried by phonons with MFPs larger than one micron~\cite{Esfarjani2011-su,Broido2007-hk}, supporting the strong phonon suppression observed in porous materials with microscale pores~\cite{Song2004-et}. Very low thermal conductivities have been measured in many nanostructures, including nanoporous materials~\cite{Yu2010-qs,Hopkins2011-ps,Tang2010-li}, nanowires~\cite{Boukai2008-hd,Hochbaum2008-ds} and thin films~\cite{Venkatasubramanian2001-uq}, corroborating the use of such material systems for thermoelectric applications. The thermoelectric figure-of-merit in semiconductors is defined as $ZT = \frac{\sigma S^2 T}{\kappa}$, where $\sigma$ is the electrical conductivity, $S$ is the Seebeck coefficient, $\kappa$ is the lattice thermal conductivity and $T$ the temperature. The numerator of ZT (the "power factor") is generally maximized at relatively high carrier concentrations, so that the average electron MFPs is of the order of a few nanometers~\cite{Qiu2015-ba}. Consequently, a properly engineered nanostructure can significantly decreases $\kappa$ with little effect on $\sigma$, yielding an increase in ZT. Despite many attempts at minimizing thermal transport in nanostructures, thermal transport optimization is still largely unexplored, primarily due to practical experimental limitations and a lack of systematic engineering approaches.

In this work, we address the latter by introducing a novel concept, the directional phonon suppression function $S(\Lambda,\Omega)$, that describes the suppression of phonons, within arbitrary geometries, with a given MFP $\Lambda$ and direction denoted by the solid angle $\Omega$. By taking into account phonons travelling both along straight lines and through multiple phonon boundary-scattering, $S(\Lambda,\Omega)$ turns out to be a powerful tool for tuning thermal transport in complex nanostructures. We employ this approach to optimize thermal transport in Si-based nanoporous materials. The heat flux is computed by means of the phonon Boltzmann Transport Equation (BTE)~\cite{Romano2016-br}. We first compute the thermal conductivity, $\kappa$, of a material system composed of a circular pores in a square lattice, finding significant heat transport degradation with respect to the bulk. Then, we use the information provided by $S(\Lambda,\Omega)$ to identify a new structure, based on rectangular pores, that exhibits $\kappa$ as small as 1 Wm$^{-1}$k$^{-1}$, well below the amorphous silicon limit. As our engineering approach can be applied to any combination of material and geometry, it paves the way to high-throughput search of ultra-low thermal conductivity materials.

\section{Results}
\begin{figure}[h!]
\begin{center}
\includegraphics[width=\columnwidth ]{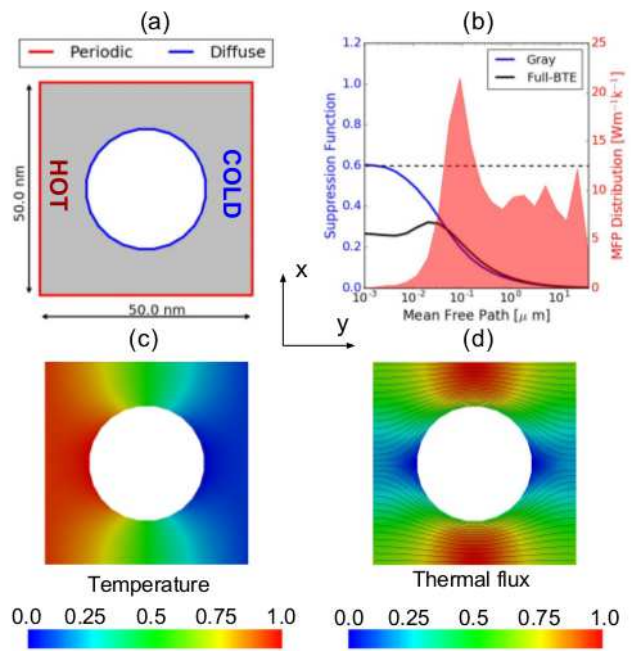}
\caption{ (a) Unit-cell comprising a single circular pore. The walls of the pores scatter phonons diffusively. Periodic boundary conditions are applied along both \textit{x}- and \textit{y}- directions. (b)  Angular average of the directional phonon suppression for different MFPs, for both the gray model and the BTE described by Eq.~\ref{Eq:1}. (c) Effective temperature distribution normalized to its maximum value. (d) Normalized magnitude of thermal flux superimposed to flux lines.}\label{Fig:1}
\end{center}
\end{figure}

In order to introduce $S(\Lambda,\Omega)$, we first compute the thermal conductivity of a nanoporous Si material with circular pores, as shown in Fig.~\ref{Fig:1}(a). Thermal transport is calculated across a unit-cell containing a single pore located at the center. We consider the unit cell sizes L = 10 nm and L = 50 nm. The diameter of the pore is set so that the porosity of the material is $\phi=0.25$. As shown in Fig.~\ref{Fig:1}(a), we apply periodic boundary conditions along both \textit{x}- and \textit{y}-directions and assume that the walls of the pores scatter phonons diffusively. A difference of temperature $\Delta T$ = 1 K is applied across the hot and cold contacts. In absence of phonon size effects, heat reduction is predicted by the Maxwell-Eucken theory and is well-described by the formula $\frac{\kappa}{\kappa_b}=\frac{1-\phi}{1+\phi}=f(\phi)$~\cite{Wang2006-by}, which is in excellent agreement with our finite-volume solver of diffusive heat conduction. Classical phonon effects are computed by means of the BTE under the relaxation time approximation~\cite{Majumdar1993-am,Romano2015-sx} 
\begin{equation}\label{Eq:1}
\Lambda \mathbf{s}(\Omega)\cdot\nabla T(\mathbf{r},\Lambda,\Omega) + T(\mathbf{r},\Lambda,\Omega) = T_L(\mathbf{r}), 
\end{equation}
where $T(\mathbf{r},\Lambda,\Omega)$ is the effective temperature of phonons with MFP $\Lambda$ and direction $\mathbf{s}(\Omega)$ and $T_L(\mathbf{r})$ is the effective lattice temperature, computed by $T_L(\mathbf{r})=\frac{1}{4\pi}\int_0^{\infty}\int_{4\pi}A(\Lambda') T(\mathbf{r},\Lambda',\Omega')d\Omega'd\Lambda'$. The weigths $A(\Lambda')$ are computed to ensure energy conservation. The input to the BTE is the cumulative function of the bulk thermal conductivity~\cite{Dames2005-ts}, computed via density functional theory~\cite{Esfarjani2011-su,Broido2007-hk}. The computed thermal conductivity of bulk Si is ~156 Wm$^{-1}$k$^{-1}$, in agreement with measurements~\cite{Glassbrenner1964-cd}. Further details about the computational approach are reported in the Methods section.

Once the mode temperature distributions are computed, $\kappa$ is obtained using Fourier's law, i.e.  $\kappa = \int_0^\infty \int_{4\pi}K(\Lambda)S(\Lambda,\Omega)d\Omega d\Lambda$, where $K(\Lambda)$ is the bulk phonon MFP distribution and 
\begin{equation}\label{Eq:2}
S(\Lambda,\Omega)=\frac{L}{\Lambda \Delta T}\int_C T(\mathbf{r},\Lambda,\Omega) \mathbf{s}(\Omega)\cdot\mathbf{n}dy.
\end{equation}
In Eq.~\ref{Eq:2}, C is the surface of the hot contact and $\mathbf{n}$ is its normal. The computed $\kappa$ are ~20 Wm$^{-1}$k$^{-1}$ and ~5 Wm$^{-1}$k$^{-1}$ for L = 10 nm, and L = 50 nm, respectively. The versor $\mathbf{s}$ can be conveniently described in terms of polar ($\varphi$) and azimuthal ($\theta$) angles, via $\mathbf{s}=\sin \theta \sin \varphi \mathbf{x} + \sin \theta \cos \varphi \mathbf{y} + \cos\theta \mathbf{z}$. The low values of thermal conductivities with respect to the bulk substantiates the presence of phonon size effects, which become stronger with smaller pore-pore distances~\cite{Romano2014-qj}. Hereafter, unless otherwise specified, we will refer to the case with L = 50 nm. We also define the Knudsen number as the ratio $Kn=\Lambda/L$. Past studies investigated nanoscale heat transport by focussing on the angular average of $S(\Lambda,\Omega)$, denoted here by $\tilde{S}(\Lambda)$. This quantity, also called "boundary scattering", is proven to be effective in MFP-reconstruction experiments and for understanding heat transport regime~\cite{Minnich2011-rs,Yang2013-tf,Minnich2015-gv}. For these reasons, we begins our investigation by analyzing $\tilde{S}(\Lambda)$.

As shown in Fig.~\ref{Fig:1}(b), $\tilde{S}(\Lambda)$ goes as $1/\Lambda$ for large Kns, as expected in ballistic transport regime. Interestingly, we observe a peak around 30 nm, below which $\tilde{S}(\Lambda)$ becomes almost flat. This result is in contrast with that obtained by employing the gray model, where heat transport is solved independently for each MFP~\cite{Romano2012-he}. This disrepancy in $\tilde{S}(\Lambda)$ for low $Kns$ arises from the fact that diffusive phonons in Eq.~\ref{Eq:1} tend to thermalize to an effective temperature (plotted in Fig.~\ref{Fig:1}(c)) that also depends on ballistic phonons. On the other hand, the effective lattice temperature within the gray formulation is simply $T_L(\mathbf{r},\Lambda)=\frac{1}{4\pi}\int_{4\pi}T(\mathbf{r},\Lambda,\Omega')d\Omega'$, which leads to a  $\tilde{S}(\Lambda)$ that is material and scale independent. The values for  $\tilde{S}(\Lambda)$, obtained by the gray model and plotted in Fig.~\ref{Fig:1}(b) show a plateau for small Kns $f(\phi)$. In fact, low-Kn phonons in the gray model thermalize only among themselves, fulfilling the requirement for recovering Fourier's law from the BTE. Finally, in Fig.~\ref{Fig:1}(d) we plot the magnitude of thermal flux, superimposed to flux lines. As indicated by the red areas, high-flux regions are in the space between the pores, as expected in the presence of strong size effects.

\begin{figure}[h!]
\begin{center}
\includegraphics[width=\columnwidth ]{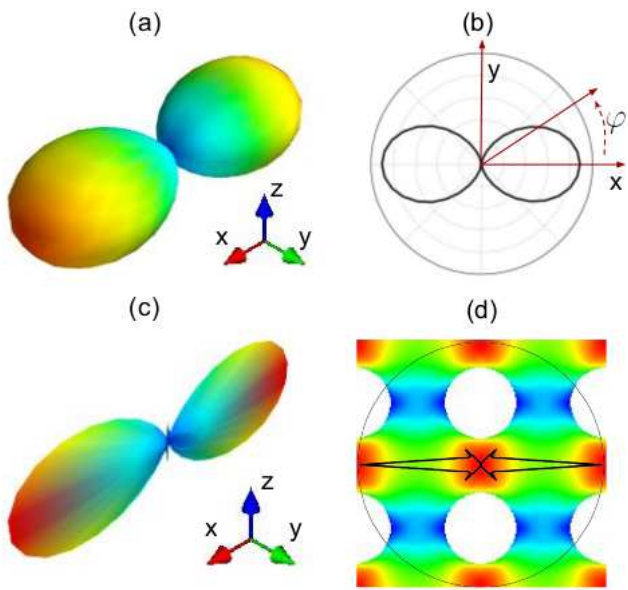}
\caption{Spherical (a) and polar (b) representations of $S(\Lambda,\Omega)$ for low Kns. The two lobes are identical because of the system symmetry. (c) Spherical and (d) polar representation of $S(\Lambda,\Omega)$ for high $Kns$. The two lobes are peaked at $\phi=0$ and $\phi=-\pi/2$. The periodicity is L= 50 nm.}\label{Fig:2}
\end{center}
\end{figure}

We now investigate the angular dependency of $S(\Lambda,\Omega)$. For very small Kns we can linearize the BTE, i.e. $\nabla T(\mathbf{r},\Lambda,\Omega)\approx \nabla T_L(\mathbf{r})$ and Eq.~\ref{Eq:1} becomes~\cite{Ziman2001-jn} 
\begin{equation}\label{Eq:3}
T(\mathbf{r},\Lambda,\Omega)\approx T_L(\mathbf{r}) - \Lambda \mathbf{s}\cdot\nabla T_L(\mathbf{r}).
\end{equation}
With no loss of generality, we use the gray model to estimate $\nabla T_L(\mathbf{r})$. According to the discussion above, the gray model reduces to Fourier's law in the diffusive regime, which is governed by the Laplace equation  $\kappa_b \nabla^2 T_L(\mathbf{r})=0$. By using the definition of porosity function, $f(\phi)$, provided above, and assuming a constant heat flux along the cold contact, we have $\nabla T_L(\mathbf{r})\approx f(\phi) \Delta T/L \mathbf{x} $. After including Eq.~\ref{Eq:3} into Eq.~\ref{Eq:2}, we obtain $ S(\Lambda,\Omega) = g(\Omega) + f(\phi)\sin^3\theta \sin^2 \varphi $. The function $g(\Omega)$ is odd in $\Omega$ and will vanish in the calculation of $\kappa$. For this reason, we will neglect it hereafter. It is interesting to note that, in this regime, $S(\Lambda,\Omega)$ is MFP-independent and that $\tilde{S}(\Lambda) = \int_{4\pi} S(\Lambda,\Omega)d\Omega = f(\phi)$, in agreement with the discussion above. The spherical and polar representations of $S(\Lambda,\Omega)$ for low Kns are plotted in Fig.~\ref{Fig:2}-a and Fig.~\ref{Fig:2}-b, respectively. We note two lobes, oriented along $\mathbf{x}$ and $-\mathbf{x}$, consistently with the fact that both the direction of the applied temperature gradient and the normal of the cold contact are aligned with $\mathbf{x}$.

For high Kns, boundary scattering becomes predominant and $S(\Lambda,\Omega)$ starts to depend strongly on the material geometry. In this regime, the LHS of Eq.~\ref{Eq:1} prevails over the RHS and it is possible to show that $S(\Lambda,\Omega)$ becomes proportional to $\sin^2 \theta$. As shown in Fig.~\ref{Fig:2}(c), $S(\Lambda,\Omega)$ is pronounced for $\phi=0$ and $\phi=-\varphi/2$, whereas rapidly vanishes for other polar angles. This trend can be explained in terms of view factor, a geometric parameter that quantifies the possibility of having direct path from the hot and cold contacts~\cite{Prasher2006-ly}. In porous materials with square pore lattices, most of the heat travels through the spaces between the pores, perpendicularly to the applied temperature gradient. The relative contribution of such paths amounts to the view factor. Fig.~\ref{Fig:2}(d) superimposes $S(\Lambda,\Omega)$ to the material geometry to better elucidate on the relationship between boundary arrangements and phonos suppression. We also note four sub-lobes in correspondence to phonons travelling along directions at 45 degrees with respect to the applied temperature gradient, constituting another set of direct paths. For all the other directions, $S(\Lambda,\Omega)$ describes Multiple Phonon-Boundary (MPB) scattering. In our case, heat transport arising from MPB is negligible, with most of the heat carried by phonons travelling through straight paths. In light of the above discussion, it is natural, therefore, to design a material where direct paths are absent.

\begin{figure}[h!]
\begin{center}
\includegraphics[width=\columnwidth ]{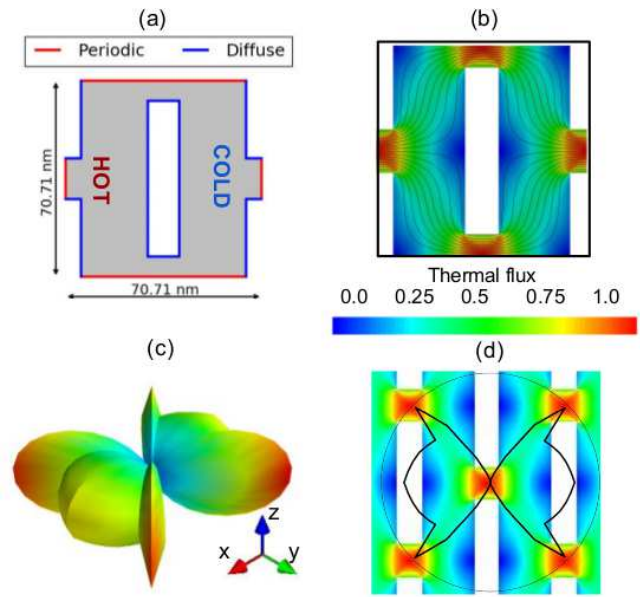}
\caption{ (a) Unit-cell of the configuration with staggered rectangular pores. (b) Normalized magnitude of thermal flux and flux lines. There are not direct paths from the cold to the hot side. (c) Spherical and d) polar representation of $S(\Lambda,\Omega)$. All thermal transport arises from multiple phonon boundary scattering.}\label{Fig:3}
\end{center}
\end{figure}
To this end, we identify a new structure with same porosity, consisting of rectangular pores in a staggered configuration, as shown in Fig.~\ref{Fig:3}(a). The periodicity is larger than the previous case because of the fixed-porosity requirement. From the flux lines plotted in Fig.~\ref{Fig:3}(b), we note that direct paths are absent thus phonons scatter multiple times before reaching the cold contact. For low Kns, $S(\Lambda,\Omega)$ is similar to that of the case with circular pores, with the size of the lobes determined by the porosity function $f(\phi)$ of the new configuration. Interestingly, for high Kns, $S(\Lambda,\Omega)$ has six preferred directions, as shown in Fig.~\ref{Fig:3}(d). The amplitudes of these peaks are significantly smaller than those in Fig.~\ref{Fig:2}(d). Remarkably, the computed $\kappa$ are ~4 Wm$^{-1}$k$^{-1}$ and ~1 Wm$^{-1}$k$^{-1}$ for L = 50nm and L = 10, respectively, almost five times smaller than their counterparts with circular pores.  We note that for the smaller size $\kappa$ is smaller than that of amorphous silicon~\cite{wada1996thermal}. Finally, as this very low thermal conductivity is obtained with no change in the porosity, electrons, which travel diffusively, will suffer moderate size effects. As a result, the configuration with staggered rectangular pores may result in a high-ZT material. \section{Conclusions}
In summary, we have introduced the directional phonon suppression function, a novel concept that captures the essence of size effects in unprecedented detail. Guided by this new quantity, we have identified a porous structure, based on rectangular pores arranged in a staggered configuration, with ultra-low thermal conductivity and relatively low porosity. Being suitable for any combinations of material and geometry, our method provides practical guidance for the development of high-efficiency thermoelectric materials and enhances our understanding of boundary-dominated heat transport. \section{Method}
The cumulative thermal conductivity requires the calculation of the three-phonon scattering time. To this end we used a 32 x 32 x 32 grid in reciprocal space and a 4 x 4 x 4 supercell. For force-constants calculations we used a 5x5x5 supercell. The isotope disorder scattering is included in the calculation. Phonon-related calculations were carried out by using ShengBTE~\cite{Li2014-fg}. Regarding electronic structure calculations, we chose a projected augmented wave (PAW) pseudopotential~\cite{Blochl1994-uj} with a plane-wave cut-off of 320 meV. Structure minimization was carried out with a 11x11x11 grid. Density functional theory calculations were performed with Quantum Espresso~\cite{Giannozzi2009-fp}. Phonon size effects were calculated by means of in-house developed parallel code. The spatial domain was discretized by means of the finite-volume method solved over an unstructured grid. The number of elements where roughly 6000. The solid angle was discretized by 48 polar angle and 12 azimuthal angles. We assumed the material be infinite along the z-direction. The MFP distribution, assumed to be isotropic, was discretized in 30 MFPs. \section{Acknowledgments}
Research supported as part of the Solid-State Solar-Thermal Energy Conversion Center (S3TEC), an Energy Frontier Research Center funded by the U.S. Department of Energy (DOE), Office of Science, Basic Energy Sciences (BES), under Award DESC0001 
\end{document}